\documentclass[pre,nofootinbib,superscriptaddress,onecolumn,preprintnumbers]{revtex4}
\usepackage{graphicx}
\usepackage{amsmath,amssymb}
\usepackage{color}

\begin{document}
\title{Symmetries of the Large Scale Structures of the Universe as a Phenomenology of a Fractal Turbulence: The Role of the Plasma Component}

\author{G. Montani}
\email{giovanni.montani@enea.it}
\affiliation{ENEA, Fusion and Nuclear Safety Department, C. R. Frascati,\\ Via E. Fermi 45,  Frascati, 00044 Roma, Italy}
\affiliation{Physics Department, ``Sapienza'' University of Rome, \\ P.le Aldo Moro 5, 00185 Roma, Italy}

\author{N. Carlevaro}
\email{nakia.carlevaro@enea.it}
\affiliation{ENEA, Fusion and Nuclear Safety Department, C. R. Frascati,\\ Via E. Fermi 45,  Frascati, 00044 Roma, Italy}

\begin{abstract}
We present a new perspective on the symmetries that govern the formation of large-scale structures across the Universe, particularly focusing on the transition from the seeds of galaxy clusters to the seeds of galaxies themselves. We address two main features of cosmological fluid dynamics pertaining to both the linear and non-linear regimes. The linear dynamics of cosmological perturbations within the Hubble horizon is characterized by the Jeans length, which separates stable configurations from unstable fluctuations due to the gravitational effect on sufficiently large (and therefore, massive enough) overdensities. On the other hand, the non-linear dynamics of the cosmological fluid is associated with a turbulent behavior once the Reynolds numbers reach a sufficiently high level. This turbulent regime leads to energy dissipation across smaller and smaller scales, resulting in a fractal distribution of eddies throughout physical space. The proposed scenario suggests that the spatial scale of eddy formation is associated with the Jeans length of various levels of fragmentation from an original large-scale structure. By focusing on the fragmentation of galaxy cluster seeds versus galaxy seeds, we arrived at a phenomenological law that links the ratio of the two structure densities to the number of galaxies in each cluster and to the Hausdorff number of the Universe matter distribution. Finally, we introduced a primordial magnetic field and studied its influence on the Jeans length dynamics. The resulting anisotropic behavior of the density contrast led us to infer that the main features of the turbulence could be reduced to a 2D Euler equation. Numerical simulations showed that the two lowest wavenumbers contained the major energy contribution of the spectrum.
\end{abstract}

\maketitle

\section{Introduction}

One of the most successful research areas in cosmology concerns the characterization of large-scale structure formation \cite{padmanabhan}. This endeavor has elucidates how primordial quantum fluctuations of the inflation field \cite{kolb,montani-primordial-cosmology,starobinski,linde} were magnified by the expansion of the Universe, becoming the seeds for structure formation. Upon re-entering the Hubble horizon, these fluctuations manifested as energy density contrasts, including scalar, vector, and tensor perturbations that are dynamically separable \cite{wein-cosm,benini2011,mukhanov}.

The linear dynamics of these perturbations—scalar, vector, and tensor—has been extensively studied in \cite{wein-cosm,kolb}. However, tracing the evolution of density contrasts and velocity fluctuations remains challenging, even within the confines of the horizon, where the Newtonian picture of the gravitational field applies. Describing structure formation on a universal scale in the non-linear regime often requires extensive $N$-body simulations to replicate the clustering processes of matter clouds soon after hydrogen recombination (see, for instance, \cite{illustris2,illustris1}).

Here, we narrowed our focus to the dynamics of the gravity--matter system, as encapsulated by three equations: the continuity equation for mass-energy density, the Euler equation predicting the velocity field evolution, and the Poisson equation governing the gravitational field in the Newtonian regime. This dynamical scheme is justified as long as we examine fluctuations in density, velocity, and gravity with wavelengths much smaller than the Hubble size, where special relativity effects remain insignificant (i.e., the velocity field remains much slower than the speed of light). This scenario often introduces the concept of the Jeans length \cite{wein-cosm,ruffini}, extending it from the linear to non-linear evolution of the fundamental fields. In other words, we considered the validity of this proposed dynamical scheme in its non-perturbative nature without reducing it to linearity when the fluctuations were small against a given background configuration.

After establishing stability conditions for linearized dynamics (i.e., introducing the Jeans length for the considered configuration), we presumed that for sufficiently large Reynolds numbers (accounting for the bulk and shear viscosity of the cosmological fluid), the dynamics naturally evolve into a turbulent regime. Here, the gravitational field minimally alters the dynamics, leading to the formation of eddies within a matter fluid associated with energy dissipation across progressively smaller scales \cite{kraichnan80,landau-v06}. Kolmogorov's postulation \cite{kraichnan80} suggests the formation of smaller eddies characterized by a constant rate of energy dissipation across various scales, filling the entire physical space. Approximately 30 years later, Mandelbrot argued \cite{mandelbrot} that eddy formation may not fill the entire spatial configuration but could be associated with a Hausdorff number less than $3$, introducing the concept of fractal turbulence. This concept successfully reproduced certain experimental features of non-linear fluid dynamics. Subsequently, Parisi and Frisch expanded the Mandelbrot idea, proposing multi-fractal turbulence where the Hausdorff number dependence on a parameter is allowed \cite{vulpiani-parisi,frisch-parisi}.

We confined our focus to simple fractal turbulence but proposed a crucial conjecture that associates the symmetries of cosmological fluid fragmentation into sub-eddies with the evolving values of the Jeans length.
Put differently, we imposed an additional constraint on fractal turbulence: the eddy scale, velocity, and mass density are related according to the Jeans formula. This natural assumption led us to derive a fundamental relationship governing the fragmentation of galaxy cluster seeds into galaxy seeds. Focusing solely on these two stages of eddy formation, which undoubtedly exist in the non-linear regime, they are likely influenced by the proposed scenario.

The primary outcome of our analysis is a constitutive formula that interlinks the values of galaxy cluster to galaxy mass densities, the number of galaxies typical within a cluster, and the Hausdorff number characterizing the fractality of these structures \cite{peebles,pietronero-book}. Such a formula can serve as a valuable tool to validate our conjecture using data sets, elucidating the extent to which the Universe exhibits a fractal structure, at least within the two main classes of observed large-scale non-linear structures.

Finally, we provide a specific cosmological implementation of the theoretical model that could describe the emergence of turbulence in the structure formation. We first included in the problem the role that a primordial magnetic field can play in influencing the non-linear regime of fluctuation growth. To this end, we analyzed the dynamical behavior of the Jeans length in the presence of a magnetic field coupled to the small plasma component of the recombined Universe. We stress how the ambipolar diffusion between the baryon neutrals and plasma contribution is responsible, at cosmological spatial scales, for a tight coupling of these two phases, which we can treat as a single fluid. According to the analyses in \cite{lattanzi-plb, montani-digioia, montani-incarbone, montani-pugliese}, we see how the Jeans length acquires non-isotropic behavior as a result of the magnetic pressure direction dependence. Since this anisotropy of the linear density contrasts (which are squeezed on the plane perpendicular to the magnetic field) is an initial condition for the non-linear problem, and since we can expect that the non-linear collapse is able to enhance such anisotropy, we arrived at studying a 2D Euler equation to describe the turbulent dynamics of the baryon fluid. Actually, the Boussinesq approximation is allowed by the high value of the plasma $\beta$-parameter. Then, we performed a numerical simulation to show how the so-called ``condensation phenomenon'' (i.e., an inverse cascade in energy) is able to limit the dominant energy content to the first two modes only (in our paradigm, clusters and galaxies). The remaining part of the energy spectrum decays with the inverse-squared wavenumbers according to a direct enstrophy cascade \cite{kraichnan67, kraichnan71, kraichnan75, kraichnan80}. We remark how a crucial role is played in the structure formation of the large-scale Universe by the presence of dark matter. However, interpreting this component of the Universe as weakly interacting particles \cite{2024arXiv240208458G,2024arXiv240208460S}, we see how no direct interaction takes place between the fractal turbulence of the baryonic matter and the cold dark skeleton of the Universe. Actually, the latter provides an external boundary condition to the dynamics of the former via the 
assignment of the dominant gravitational contribution. In other words, the backreaction of the turbulent baryonic matter on the dark energy component can be neglected with a good degree of accuracy. The assignment of a realistic background gravitational field coming from the dark matter contribution on which the present turbulence scheme had to be evolved is beyond the scope of the present analysis, though it remains an important issue for a more 
predictive investigation of the proposed picture. We account for the presence of dark matter by including it in the background Universe density, which interacts with the gravitational field.

This manuscript is structured as follows. In Section \ref{sec2}, we provide the theoretical framework and the dynamical system underlying the proposed idea. The linear stability of the cosmological fluid is discussed and the paradigm of the Jeans fragmentation is outlined. In Section \ref{sec3}, we implement the fractal turbulence scenario for the cosmological fluid non-linear dynamics. The so-called $\beta$-model is applied when scale lengths and velocity fields are related by the Jeans formula. A phenomenological law is determined when considering the fragmentation of galaxy cluster seeds into galaxy seeds. This result is shown to be a valuable tool to investigate the possible Universe fractality. In Section \ref{sec4}, we propose a cosmological implementation of the fractal turbulence scenario by considering the role of the primordial magnetic field and studying the evolution of the Jeans length. A 2D reduction in the dynamics was argued and numerical simulations of the turbulent energy spectrum were performed. Concluding remarks follow.

\section{The Jeans Length for a Viscous Cosmological Fluid}\label{sec2}

We now investigate the gravitational stability of a cosmological fluid taken well inside the Hubble horizon of the considered stage of evolution. In this scenario, the gravitational field can be properly treated on a Newtonian footing and a fluid representation of the cosmological medium is appropriate. In particular, here we take into account the shear and bulk viscosity effects, which are justified by the inhomogeneity of the perturbations and by the rapid expansion of the nearly recombined Universe, respectively.
In fact, on the one hand, the presence of density inhomogeneities and of their associated velocity fields makes the resulting picture compatible with the friction of adjacent fluid layer and, on the other hand, the Universe expansion hinders 
the thermodynamic equilibrium, requiring the introduction of bulk 
dissipation to account for small deviations from such an equilibrium. 
Clearly, these viscosity contributions are expected to be very small and their relevance increases with the wavenumber of the perturbations \cite{carlevaro-mpla,wein-cosm,montani-digioia}, but it is just in correspondence with high values of the Reynolds number (which are surely guaranteed for a weak viscous fluid) that the turbulence phenomena discussed below take place.

The dynamics of the cosmological fluid inside the Hubble horizon is governed by the continuity equation (ensuring the mass conservation of the system): 
\begin{equation}
	\partial_t\rho + \boldsymbol{\nabla}\cdot (\rho \textbf{v}) = 0
	\, ,
	\label{jea1}
\end{equation}
by the Euler equation (guaranteeing the momentum conservation of the system):
\begin{equation}
	\rho \left( \partial _t\textbf{v}  + \textbf{v}\cdot \boldsymbol{\nabla}\textbf{v}\right) = - \boldsymbol{\nabla}p - \rho \boldsymbol{\nabla}\Phi + \eta\Delta \textbf{v} + \Big( \frac{1}{3}\eta + \iota \Big) \boldsymbol{\nabla}(\boldsymbol{\nabla}\cdot \textbf{v}) 
	\, ,
	\label{jea2}
\end{equation}
and, eventually, by the Poisson equation (describing the self-gravitation of the fluid):
\begin{equation}
	\Delta \Phi = 4\pi G \rho
	\, .
	\label{jea3}
\end{equation}
Above, 
$\rho$ denotes the mass energy density, $p$ denotes the pressure, $\Phi$ denotes the gravitational potential, and $G$ denotes Newton's constant. Furthermore, $\eta$ and $\iota$ stand for the shear and bulk viscosity coefficients, respectively. We also observe that the following relation holds:
\begin{equation}
	\boldsymbol{\nabla}p = \frac{dp}{d\rho}\boldsymbol{\nabla}\rho \equiv v_s^2 \boldsymbol{\nabla}\rho
	\, , 
	\label{jea4}
\end{equation}
where we assumed the validity of a barotropic equation of state $p=p(\rho)$ and $v_s$ denotes the sound speed. Introducing the specific (or kinematical) coefficients 
$\nu \equiv \eta  /\rho$ and $\xi \equiv \iota /\rho$, we can now rewrite Equation (\ref{jea2}) in the form
\begin{equation}
	\partial _t\textbf{v} + \textbf{v}\cdot \boldsymbol{\nabla}\textbf{v} = - \frac{v_s^2}{\rho} \boldsymbol{\nabla}\rho - \boldsymbol{\nabla}\Phi + 
	\nu \Delta \textbf{v} + \Big(\frac{1}{3}\nu + \xi \Big) \boldsymbol{\nabla}\left( \boldsymbol{\nabla}\cdot \textbf{v}\right)
	\, . 
	\label{jea5}
\end{equation}

The equation above 
is crucial for the study of the fractal turbulence in the next section in view of its scale invariance. Actually, the major ($\simeq$80\%) non-relativistic energy density contribution comes from the dark matter component, which, being (reliably) constituted by weakly interacting particles, is essentially pressureless. However, we included a non-zero pressure term in the dynamics of the cosmological fluid because the baryon matter component ($\simeq$20\%) is strongly
coupled to the radiation, not only when the Universe is a plasma, but also after the recombination era. A non-negligible radiation pressure also existing
in the recombined Universe is a consequence of the very large photon/baryon
ratio ($\simeq$$10^9$) such that the photon--baryon scattering generating and a non-zero pressure term has to be included in the cosmological dynamics up to
a redshift $\simeq$100 \cite{lattanzi-plb,wein-cosm}.

\subsection{Linear Stability}
Here, we study the linear stability of the dynamical system of Equations (\ref{jea1})--(\ref{jea3}) around a background configuration characterized by a constant and uniform mass energy density $\rho_0$; a zero gravitational potential (since we consider the fluid homogeneity on a spatial scale much greater than the typical perturbation wavelength); and, eventually, a null velocity field. Hence, we can decompose the fluid density as $\rho = \rho_0 + \delta \rho$ 
(here $|\delta\rho|\ll\rho_0$), and we introduce the density contrast $\delta \equiv \delta \rho /\rho_0$. 
For the gravitational potential and the velocity field, we retain the notation above for the perturbations 
(i.e., $\Phi$ and $\textbf{v}$) since they have a null background component.

If we linearize the system around the considered background, we obtain the following equations:
\begin{align}
	\partial_t \delta = - \boldsymbol{\nabla}\cdot \textbf{v} 
	\, ,
	\label{jea6}\\
	\partial_t\textbf{v} = - v_s^2 \boldsymbol{\nabla}\delta - \boldsymbol{\nabla}\Phi 
	+ \nu \Delta \textbf{v} + 
	\Big( \frac{1}{3}\nu + \xi\Big)\boldsymbol{\nabla}\left(\boldsymbol{\nabla}\cdot \textbf{v}\right) 
	\, ,
	\label{jea7}\\
	\Delta \Phi = 4\pi G\rho_0\delta
	\,.\label{jea8}
\end{align}
Taking 
the divergence of Equation (\ref{jea7}) and using both Equations (\ref{jea6}) and (\ref{jea8}), we obtain the following equation for the density contrast:
\begin{equation}
	\partial^2_t\delta = 
	v_s^2\Delta \delta + 4\pi G\rho_0\delta + \mu \partial_t\Delta \delta
\, ,
	\label{jea9}
\end{equation} 
where $\mu \equiv 4\nu /3 + \xi$. 
For this equation, taking the plane wave solution 
$\delta \sim e^{i(\textbf{k}\cdot \textbf{x} - \omega t)}$ 
provides a dispersion relation of the form
\begin{equation}
\omega^2 - k^2v_s^2 + 4\pi G\rho_0 + 
i\omega \mu k^2 = 0
\, 	\label{jea10}
\end{equation}
in the Fourier space of frequencies $\omega$ and wavenumbers $k=|\textbf{k}|$ (the direction of the wavevector is not involved for a homogeneous and isotropic background). The associated wavelength is defined as $\lambda=2\pi/k$.

To extract the physical content of this dispersion relation, we set $\omega = \omega_r + i\gamma$ so that its imaginary part provides $\gamma = - \mu k^2/2$, which ensures a net damping due to the viscosity terms. However, in what follows, we are interested in the so-called ``inertial region'', where the viscous dissipation does not prevent the development of 
turbulent behavior; therefore, we can neglect, with good physical significance, the terms of Equation (\ref{jea10}) 
proportional to $\mu$. The resulting system stability is evaluated from the condition $\omega^2\geqslant0$ and it is ensured by the standard Jeans condition \cite{jeans,ruffini} 
\begin{equation}
\lambda \leqslant \lambda _J \equiv \sqrt{\frac{\pi v_s^2}{G\rho_0}} 
	\, , 
	\label{jea11}
\end{equation}
where $\lambda_J$ denotes the Jeans length. A detailed analysis of the dispersion relation is provided in Appendix A (see also \cite{carlevaro-mpla}).

\subsection{Jeans Fragmentation Mechanism}

The formula of the Jeans length, when analyzed in the context of non-linear stages of evolution, suggests the idea that a possible mechanism of fragmentation of the cosmological fluid in substructures could have taken place in the formation of large-scale structures in the Universe \cite{ruffini-fragmentation}. This idea also provides the theoretical framework 
for proposing fractal features of the matter distribution across the Universe \cite{labini-1998,calzeti1,calzetti2}.

According to Equation (\ref{jea11}), we assumed that at given time $t_1$, the value of the Jeans length is $\lambda_{J1}$. Then, all the scales $\lambda > \lambda_{J1}$ can start growing and, consequently, at least one of them will achieve the non-linear regime in which the density contrast is greater than unity. In such a regime, the overdensity starts to gravitationally collapse and it increases its density. At a given instant $t_2>t_1$, the overdensity 
size becomes comparable with the new value of the Jeans length $\lambda_{J2} < \lambda_{J1}$, i.e., 
\begin{equation}
\lambda_{J2} = \sqrt{\frac{\pi v_s^2}{G \rho_1}} <\lambda_{J1}
\, , \label{jea12}
\end{equation}
where we assumed $\rho_2>\rho_1$, while we retained an almost constant sound speed value. 

This scheme can be iterated to produce the fragmentation of the original homogeneous patch in a series of homogeneous substructures of higher density. However, no concrete proof exists that this scenario could have been a real mechanism in the large-scale structure formation across the early Universe. In what follows, we concentrate our attention to a specific step of the structure formation, i.e., the passage from clusters of galaxies to single galaxies, which is the only clear representation of structures vs. substructures that occur in the non-linear regime today. 

The idea here proposed does not rely only on the gravitational physics to justify the fragmentation of galaxy clusters into isolated substructures, but we interpreted this process as a consequence of the non-linear feature of the fluid dynamics in its full nature. More specifically, we assumed that the fragmentation should be interpreted as a phenomenon of the fractal turbulence characterizing the fluid dynamics in the inertial region, where the value of the Reynolds number is sufficiently large \cite{mandelbrot,vulpiani-parisi,frisch-parisi}. In other words, we considered the non-linear structure to be identified with the seed of a galaxy cluster as a large turbulent eddy that is subjected to a spontaneous fragmentation into a class of smaller eddies because of the energy transport from larger to smaller spatial scales. The physical motivation for such fragmentation is, therefore, the tendency of the homogeneous patches to gravitationally collapse according to the driving process of a direct energy cascade across turbulent eddies.

\section{Fractal Turbulence}\label{sec3}

In this section, we develop some basic considerations regarding the fractal turbulence referred to in the dynamical system of Equations (\ref{jea1}), (\ref{jea3}), and (\ref{jea5}) in order to characterize, via the simple model discussed in \cite{vulpiani-parisi}, the turbulent eddy fragmentation. Then, identifying the eddy spatial scale with the Jeans length for the structure formation, we define a basic scale law linking the density contrast of galaxies over galaxy clusters to the number of galaxies in each cluster (in what follows, we limit our attention to a single non-linear fragmentation step).

\subsection{Scaling Law of the Dynamics}

Conversely to the analysis presented in \cite{vulpiani-parisi}, the considered dynamical equations contain the bulk viscosity term and, more importantly, the additional gravitational interaction represented by the gradients of the potential $\Phi$ in Equations (\ref{jea3}) and (\ref{jea5}). It is easy to see that Equation (\ref{jea5}) is scale invariant by a factor $\theta^{2h-1}$ ($\theta$ and $h$ are real numbers) under the following re-parameterization:
\begin{equation}
\textbf{x}\rightarrow \theta \textbf{x} \, ,\quad
t\rightarrow \theta^{1-h}t\, ,\quad 
\textbf{v}\rightarrow \theta^{h}\textbf{v}\, ,\quad
\Phi\rightarrow \theta^{2h}\Phi 
	\, , 
	\label{jea13}
\end{equation}
and the constants $v_s$, $\nu$, and $\xi$ are scaled according to their dimensionality. Furthermore, from Equation (\ref{jea3}), we obtain the scaling law for the density in the form
\begin{equation}
\rho \rightarrow \theta^{2h-2}\rho
\, , \label{jea14}
\end{equation}
which implies the following scaling for the mass:
\begin{equation}
M\equiv \int_V\rho\; dV \rightarrow \theta^{2h+1}M
	\, ,
	\label{jea15}
\end{equation}
where $V$ denotes a generic space volume. Clearly, as it must be, Equation (\ref{jea14}) is consistent with the Jeans length definition in Equation (\ref{jea11}).

The energy dissipated via the the shear viscosity (the situation is similar in its estimates 
for the bulk contribution) is given by
\begin{equation}
	\epsilon \sim \nu\left( 
	\boldsymbol{\nabla}\textbf{v}\right)^2
	\, , 
	\label{jea16}
\end{equation}
which implies the well-known scaling law
\begin{equation}
	\epsilon \rightarrow \theta^{3h-1}\epsilon
	\, .
	\label{jea17}
\end{equation}
According 
to the conjecture of Kolmogorov \cite{kraichnan80} stating that the quantity $\epsilon$ has to be constant across the spatial scales, we make the corresponding choice $h=1/3$. As discussed in \cite{vulpiani-parisi}, this scenario implies that the velocity gradients are characterized by singular points across space since the following relation holds:
\begin{equation}
	\lim_{\theta\rightarrow 0} \frac{\!\!\!\!\!\!\mid \Delta \textbf{v}\mid}{\;\mid \Delta \textbf{x}\mid^{1/3}} \neq 0
	\, .
	\label{jea18}
\end{equation}

However, Kolmogorov assumed that these singular points, which are de facto turbulent eddies, continuously fill the space, while Mandelbrot \cite{mandelbrot} proposed the idea that such singular points are distributed according to a fractal structure. In particular, Mandelbrot postulated that the probability of having a singular point in space is of the form $P_{\theta}\sim \theta^{3 - D}$, where $D$ is a positive constant $0< D \leqslant 3$, which is dubbed the ``Hausdorf number''. Then, this idea was generalized in \cite{frisch-parisi} to the so-called ``multi-fractal turbulence'', i.e., in order to account for the behavior of the velocity correlation functions, it was postulated that the Haushdorf number must become a function of the scaling exponent $D=D(h)$.

\subsection{Generalized $\beta$-Model}

We now implement the fractal turbulence scenario \cite{mandelbrot} to the cosmological fluid dynamics by identifying the sub-eddy formation with the Jeans fragmentation scales. In particular, we adopt the so-called $\beta$-model, as discussed in \cite{vulpiani-parisi} and slightly generalized to the present context. 

Starting from an eddy with a spatial scale $\ell_0$, we consider the following fragmentation law in a series of sub-eddies:
\begin{equation}
\ell_n = \ell_0\left(\frac{1}{N}\right)^n	\, , 
	\label{jea19}
\end{equation}
where $N$ is a fixed integer number (in \cite{vulpiani-parisi}, it was taken to be equal to $2$) and $n=1,2,3,\dots$ According to Equation (\ref{jea16}), the energy dissipation behaves as $\epsilon \sim v^3/\ell$ such that we obtain the relation
\begin{equation}
\epsilon_n \sim \beta^n \frac{v_n^3}{\ell_n}\, , 
	\label{jea20}
\end{equation}
where $v_n$ is the velocity scale associated with the eddy of length scale $\ell_n$ and $\beta < 1$ is a factor that accounts for the fractal nature of the fragmentation, i.e., it ensures that the eddy formation does not fill the whole physical space. It is natural to take the $\beta$-parameter equal to the probability to have a specific spatial scale, that is, $\beta^n = (\ell_n/\ell_0)^{3-D} = N^{(D-3)n}$. Furthermore, since the energy dissipation is assumed to be scale independent ($\epsilon_n\to\epsilon$), we easily arrive at the following expression for the velocity scale \cite{vulpiani-parisi}: 
\begin{equation}
	\frac{v_n}{v_0} = \left( \frac{\ell_n}{\ell_0}\right)^{(D-2)/3} = N^{(2-D)n/3}
	\, . 
	\label{jea21}
\end{equation}

Henceforth, we focus our attention on the first step of eddy fragmentation, i.e., setting $n=1$ in the formula above. Furthermore, we implement this scenario in the passages from galaxy cluster seeds ($n=0$) and galaxy seeds ($n=1$). According to this idea, we identify the fragmentation scales with the Jeans length values and the eddy velocity with the sound velocity. Then, from Equation (\ref{jea11}), we obtain
\begin{equation}
	\frac{\rho^{g}}{\rho^{c}} = 
	\left( \frac{\lambda_J^{c}}{\lambda_J^g}\frac{v_s^g}{v_s^{c}}\right)^2
	\, , 
	\label{jea22}
\end{equation}
where the labels $(c)$ and $(g)$ indicate galaxy clusters and galaxies, respectively. Combining Equations (\ref{jea21}) and (\ref{jea22}), we arrive at the fundamental relation between the ratio of the galaxy over the galaxy cluster energy density and the number of galaxies within a single cluster, i.e.,
\begin{equation}
	\frac{\rho^g}{\rho^{c}}=N^{2(5-D)/3}\, .
	\label{jea23}
\end{equation}

\begin{figure}[ht!]
\centering
\includegraphics[width=10cm]{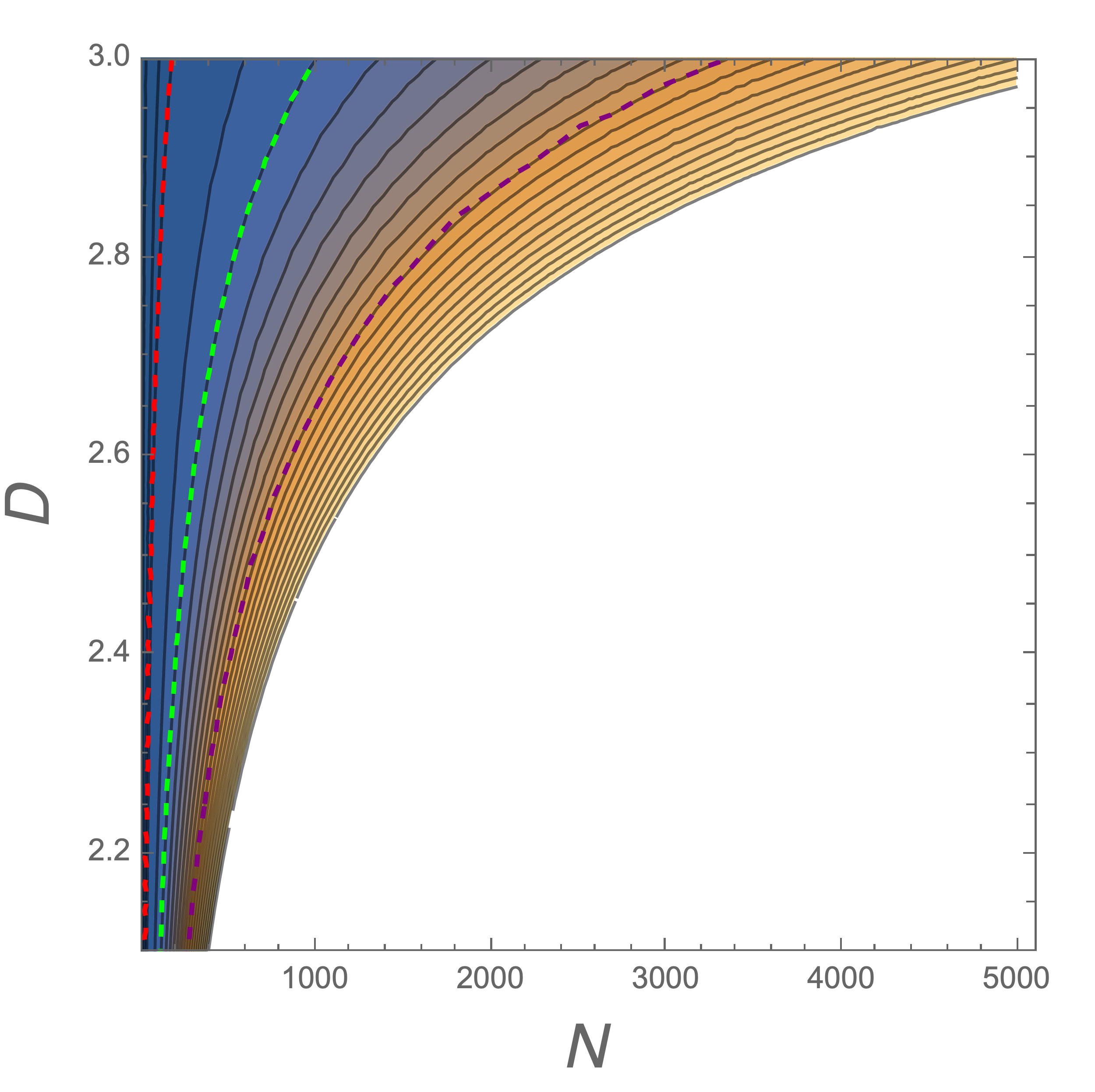}
\caption{Contour plot of $\rho^g/\rho^{c}$ from Equation (\ref{jea23}) as a function of $N$ and $D$. The white excluded region represents $\rho^g/\rho^{c}>10^5$. Colored dashed lines correspond to $\rho^g/\rho^{c}=1000$ (red), 10,000 (green), and 50,000 (purple).
\label{fig1}}
\end{figure}

This 
formula constitutes a phenomenological tool to investigate data on the large-scale matter distribution across the Universe in order to determine the average Hausdorf number of the structure formation process. In this respect, let us consider the following estimate~\cite{peebles} for the average values of the mass densities: $\rho^g \in(10^{-24},10^{-22})$ g/cm$^3$ and $\rho^{c} \in(10^{-27},10^{-25})$ g/cm$^3$. We, thus, obtain $\rho^g/\rho^{c} \in(10,10^{5})$, and in Figure \ref{fig1}, we plot the contours of the density ratio from Equation (\ref{jea23}). As an example, let us consider a galaxy cluster with $\rho^g/\rho^{c}=5\times10^4$: if the number of the contained galaxies is $N=2000$, the fractal clustering index is $D\simeq2.86$, while reducing the galaxy number to $N=1000$ it results in $D\simeq2.65.$

In general, we see that the fractality of the structure formation
emerges in correspondence with a given density ratio
in the region of smaller numbers
of constituents (galaxies) with respect to larger ones, where $D\simeq 3$. This result offers an interesting tool to search across the Universe
galaxy distributions with dimensionality $D<3$, especially in those
clusters that are less populated.
The validity of this theoretical conjecture, as expressed by the predictivity
of Equation (\ref{jea23}), can be tested only via a systematic and statistical
analysis of the large-scale matter distribution, which will be allowed by
the James Webb Space Telescope \cite{jwst} and Euclid \cite{euclid}.

We conclude by stressing that we chose to implement the fractal turbulence scenario to clusters of galaxies and galaxies only because the application of the proposed theoretical scheme requires fully developed non-linear fluid dynamics and these two structures of the Universe are the only ones in a surely non-linear regime among the most important spatial scales for the matter distribution.

\section{Cosmological Implementation of the Model}\label{sec4}
We now set a possible reference 
scenario in which the idea that the 
fractal turbulence triggers the galaxy formation from galaxy clusters can be properly assessed. We considered a small background magnetic 
field, as analyzed for the linear dynamics in \cite{lattanzi-plb}, in order to elucidate the possibility that the relevant turbulent behavior of the fluid is mainly reduced to a 2D Euler equation. A numerical analysis was then performed in the limit when the so-called ``condensation phenomenon'' \cite{kraichnan80} dominates the turbulence spectrum \cite{biskamp-book}.

\subsection{The Evolution of the Jeans Length}
The initial conditions for the linear perturbation evolution are provided by the scale-invariant spectrum 
generated during the de Sitter phase of the inflationary Universe~\cite{kolb,wein-cosm,montani-primordial-cosmology}, according to which the perturbations re-enter the Hubble horizon 
(in the radiation or matter-dominated era of the Universe), all with the same amplitude, independently of their~wavelength. 

During the radiation-dominated Universe, the equation of state of the cosmological fluid is $p_r= \rho_r/3$ 
(the suffix $r$ refers to the radiation component) and the sound 
velocity remains correspondingly constant and equal to the value $c/\sqrt{3}$. Since the energy density decreases with decreasing redshift $z$ as $\rho_r\propto (1 + z)^4$, clearly, the Jeans length 
during the radiation-dominated era increases as $\lambda_{Jr}\propto (1 + z)^{-2}$, according to Equation (\ref{jea11}). Apart from irrelevant numerical factors, such behavior also remains valid in the 
case of an expanding Universe well inside the Hubble scale \cite{montani-primordial-cosmology}. This picture for the Jeans length evolution also holds after the equivalence age and up to the recombination. 
This is due to the fact that as far as the Universe is in the state of a plasma, the baryon--photon scattering 
is efficient, as in the radiation-dominated era. The only difference is that now, the energy density is dominated 
by the matter and the Jeans length 
increases as $(1 + z)^{-3/2}$.

The situation drastically changes when 
the hydrogen recombines, becoming transparent to photons. However, also 
after the recombination, the radiation pressure significantly affects the 
cosmological fluid dynamics simply because, due to a photon/baryon ratio 
of about $10^9$, a photon scattering on a baryon strongly survives from the redshift value $z\simeq 1100$ up to 
$z\simeq 100$. In this redshift interval, while the pressure is still 
$p_r\simeq \rho_r/3$, the Universe 
density is dominated by the matter contribution, i.e., $\rho_0\simeq\rho_m$, where the suffix $m$ refers to the matter component. Thus, an approximated equation of state for the Universe in this stage is $p_r\simeq\rho_m K_B T_b/m_b\propto\rho_m^{4/3}$ (here, $T_b$ is the baryon temperature, $m_b$ is the baryon mass, and $K_B$ is the Boltzmann constant) \cite{wein-grav}, and for the corresponding sound velocity, we~obtain
\begin{equation}
	v_s^2(100<z<1100) \simeq 
	\frac{4}{3}\frac{p_r}{\rho_m}\simeq \frac{4}{3}\frac{K_B T_b}{m_b}\propto	(1 + z)	\, .
	\label{jea1vs}
\end{equation}

This 
formula states that the sound velocity has a jump by a factor about $10^{5}$, which agrees with 
the recombination value of the matter density.
As a result, we see that near the recombination era, the Jeans length has its maximum \cite{lattanzi-plb}. Thus, to ensure that the Jeans mechanism of fragmentation can concern 
galaxy cluster and galaxies, we 
have to require that their present day mass (a constant of the structure evolution) is contained in a Jeans sphere of volume $\lambda_J^3$, i.e.,
\begin{equation}
	M_{c,g} < \rho_m^{rec}({\lambda_J^{rec}})^3 \simeq \frac{c^2}{G}\lambda_J^{rec}
	\, , 
	\label{jea12B}
\end{equation}
which can be easily restated as follows:
\begin{equation}
	R_S^{c,g} \equiv \frac{2GM_{c,g}}{c^2} < 2 \lambda_J^{rec}\simeq 10^{23}\textrm{cm}
	\, .
	\label{jea13B}
\end{equation}
Above, 
the label $rec$ refers to the recombination era and $R_S$ denotes the Schwarzschild radius of the considered objects.
The condition above is clearly satisfied since the Schwarzschild radius of the Sun is about $3\times 10^5$ cm, while a galaxy contains about $10^{12}$ solar masses and a galaxy cluster 
is constituted by $10^2$ to 
$10^4$ galaxies. Thus, the estimate above clearly 
allows for implementing the concept of 
Jeans fragmentation and then turbulent dynamics at the spatial scales of 
galaxies and galaxy clusters since they are already able to increase 
in the linear regime.

\subsection{The Role of the Magnetic Field}

After the recombination, say from 
$z\simeq 1100$ up to $z\simeq 10$, where the large scale structure formation takes place, the cosmological fluid is 
constituted by three main matter components: the dark matter, the neutral baryonic matter, and the 
plasma baryonic relic (actually a fraction of the primordial plasma, i.e., $x_e\simeq2.5\times10^{-4}$, cannot recombine due to the expansion rate \cite{kolb,novos}). Although the plasma component is 
very small, particularly 
compared with the radiation contribution, its presence has a relevant dynamical implication. As discussed in~\cite{lattanzi-plb}, this important role of the plasma is due to the intense ambipolar diffusion term~\cite{mestel-amb,shu-amb,banerjee-amb,li-amb}, which significantly couples the ionized and neutral components at large enough cosmological scales. 
This feature becomes dynamically important if we postulate the presence of 
a primordial magnetic field according to well-grounded theoretical proposals \cite{barrow-rept} and the observation of an appreciable field at the present galaxy scale \cite{magnetic-galactic}. 
According to the constraints coming from the microwave background radiation \cite{2023MNRAS.518.3675D,Planck:2018vyg,giovannini}, the strength of the primordial magnetic field at the 
recombination age cannot exceed the 
value of about $10^{-9}$ G and we recall that its intensity scales as 
$(1 + z)^2$. The important point here is that the magnetic field influences the linear and non-linear dynamics of the fluctuations in the plasma component and, hence, the dynamics of 
the baryonic fluid. 
However, before discussing how the 
linear dynamics of the perturbation are influenced by the magnetic field and 
which implications can come out on the turbulence dynamics, a brief digression must be dedicated to the role of dark matter. 

We aligned our point of view to the main scientific stream that interprets dark matter in terms of weakly interacting particles, leaving a very small trace on the last scattered photons, but providing the dominant gravitational skeleton for the structure formation. At the recombination age, the 
analysis of the phenomena mainly characterizing the photon--matter interaction \cite{padmanabhan} fixes the baryonic density contrast as $\delta \rho_b/\rho_b\simeq 10^{-3}$. Since this linear density contrast evolves as $(1+z)^{-1}$, today, it would be of order unity and 
the non-linear structure formation process could not have taken place. 
Thus, we are led to postulate that at the recombination, the density contrast of the dark matter fluctuations were about $20$~times greater than that 
of the baryonic component.

In principle, the dark matter contribution to the Universe density could be treated as a pressureless fluid, although its weak collisionality could suggest the necessity of a kinetic description of its dynamics. However, 
for what concerns the problem of the 
baryonic large-scale structures (say, the visible component of cluster of galaxies), the dark matter presence can be regarded as an assigned background gravitational component on which the baryonic (neutral and non-neutral) fluid evolves. The detailed account of this gravitational interaction between dark and baryon matter could be clearly accommodated in the proposed dynamical
scenario by means of the system Equations (\ref{jea1})--(\ref{jea3}) as soon as we consider a more realistic perturbation scheme in which the background gravitational field is inhomogeneous and provided by a complementary study of the dark matter gravitational instability. 
This perspective can be reliable pursued, but it has some challenging aspects, especially regarding the 
necessity of numerically demanding $N$-body simulations to describe the highly non-linear matter clustering \cite{illustris2,illustris1}. To obtain a physical insight of the 
influence that a magnetic field can have on the presently proposed picture, we followed the same approximation adopted in \cite{lattanzi-plb}, where the concept of the Jeans length is recovered in the presence of both the Universe's expansion and a constant background magnetic field $\textbf{B}_0$. Thus, when we consider the propagation of the baryon fluctuations, the background quantities are calculated using the baryonic matter energy density only, while when the gravitational field enters the problem, it interacts with the full Universe energy density $\rho_0$, including the dark matter term.

For a detailed discussion of the linear perturbation dynamics in the case of a magnetized cosmological fluid, we refer the reader to \cite{barrow-rept}. However, coherently with the discussion in Section \ref{sec2}, below we discuss the main physical effects, limiting our attention to a static background, which is surely a good approximation for the case of short perturbation wavelengths (i.e., much less than the Hubble horizon). According to the considerations above, if we treat the (ionized and neutral) baryonic matter as a single magnetized fluid, following a linear perturbation analysis similar to the one in Section \ref{sec2} (see Appendix B), the resulting solution of the dispersion relation takes the following form:
\begin{equation}
\omega^2=\omega^2_{\pm} = \frac{k^2}{2}
	\left[ \tilde{v}^2 + v_A^2 \pm 
	\sqrt{\left( \tilde{v}^2 + v_A^2\right)^2 - 4\tilde{v}^2v_A^2\cos^2\theta}\right]
	\, , 
	\label{jea1x}
\end{equation}
where $v_A$ denotes the background Alfv\'en velocity ($v_A^2\equiv B_0^2/4\pi \rho_b$), $\tilde{v}^2\equiv v_s^2 - 4\pi G \rho_0/k^2$, and $\cos \theta \equiv 
\hat{\textbf{B}}_0\cdot\hat{\textbf{k}}$. We stress how, according to \cite{montani-incarbone}, we calculate the Alfv\'en velocity with the baryon matter density only, while for $\tilde{v}^2$, we used the full Universe matter energy density.

We separately analyzed the two limiting cases $\theta = \pi/2$, i.e., the wavevector lying in the plane perpendicular to the magnetic field, and $\theta = 0$, i.e., when the perturbations propagate along the magnetic field direction. When $\theta = \pi /2$, the solution with the  $(-)$ sign in Equation (\ref{jea1x}) provides $\omega_- = 0$, i.e., we deal with a trivial redefinition of the background. In the opposite case, when we 
retain the $(+)$ sign, it is easy to recognize that the perturbations grow when $k^2< 4\pi G \rho_0/(v_s^2 + v_A^2)$, which corresponds to restating the Jeans length as follows:
\begin{equation}
\lambda_J \equiv \sqrt{\frac{\pi\left( v_s^2 + v_A^2\right)}{G\rho_0}}\;.
\label{jeax2}
\end{equation}

This 
result has a clear physical explanation: in the fast magnetoacoustic branch \cite{biskamp-book2}, the magnetic field strength provides a pressure contribution, i.e., 
$\textbf{B}_0\cdot\delta\textbf{B}/4\pi$, with $\delta\textbf{B}$ being the field perturbation, which supports the thermal pressure and then increases the Jeans length~value. 

When we consider $\theta = 0$, we immediately obtain
\begin{equation}
	\omega_{\pm}^2 = \frac{k^2}{2}\left( \tilde{v}^2 + v_A^2 \pm 
	|\tilde{v}^2 - v_A^2|\right)\;,
	\label{jeax4}
\end{equation}
and from a careful inspection of both the sign cases, we simply obtain that the gravitational instability takes place for perturbations with a wavelength larger than the standard Jeans length of Equation (\ref{jea11}) (propagating Alfv\'en modes are also available). This feature has a natural interpretation in terms of the geometry of the problem. In fact, the perturbed magnetic field is naturally divergenceless, and then $\delta\textbf{B}\cdot \textbf{k} = 0$; since $\textbf{k}$ is aligned with $\textbf{B}_0$, we must also have $\textbf{B}_0\cdot\delta\textbf{B}=0$. In other words, we are dealing with a vanishing perturbed magnetic pressure and the Jeans length remains unaffected by the non-zero magnetic shear contribution. 

In the oblique case $0<\cos\theta<1$, the situation is intermediate, as discussed in detail in \cite{montani-incarbone} (see also \cite{montani-pugliese}), and anisotropic behavior of the linear perturbation is introduced. In fact, an initial spherical overdensity in space tends, as the system evolves, to squeeze on the plane perpendicular to the background magnetic field. Thus, we are led to infer that when the tight coupling between the neutral and ionized baryon component of the Universe is considered, the non-linear evolution of the density contrast, which has the linear evolution as the initial and boundary conditions, could exhibit a certain axial symmetry characterized by a tendency to form squeezed structures (see \cite{germani} for the derivation of a similar scenario for a strong field context).

Furthermore, the plasma configuration that we are considering has a high value for the $\beta$-parameter, i.e.,
\begin{equation}
	\beta \equiv \frac{8\pi p_r}{B_0^2} \simeq 0.2\frac{c^2}{v_A^2\mid_{t=t_{rec}}} \gg 1 \, .
	\label{jea1x5}
\end{equation}
This 
value is $z$-independent, and due to the large ambipolar diffusion at large scales, we can apply the Boussinesq approximation to the baryonic fluid \cite{balbus}. Thus, we can assume 
that the velocity field is dominated by its divergenceless component. We can finally write the fluid velocity field as 
\begin{equation}
	v_x\simeq -\partial_y\psi \, ,\qquad
 v_y\simeq \partial_x \psi \, ,\qquad 
	v_z\simeq 0
	\, , 
	\label{jea1x6}
\end{equation}
where $\psi(x,y)$ is the stream function \cite{landau-v06}, and we have assumed, without loss of generality, that the constant and uniform magnetic field $\textbf{B}_0$ is along the $z$-direction. 

Now, by means of the velocity field above and taking the curl of Equation (\ref{jea5}), we easily arrive at the following viscous Euler equation:
\begin{equation}
	\partial_t\Delta\psi + 
	\partial_x\psi\partial_y\Delta \psi - \partial_y\psi\partial_x\Delta \psi = \nu\Delta^2\psi
	\, , 
	\label{jea1x7}
\end{equation}
where $\Delta$ denotes the 2D Laplacian operator. We stress that $(\boldsymbol{\nabla}\times \textbf{v})_z = \Delta \psi$ is the vorticity field, and the associated enstrophy cascade \cite{kraichnan80} is responsible for the fragmentation of the turbulent eddies at smaller and smaller scales. The presence of vorticity at the galaxy spatial scale suggests that in proto-galaxies, a certain amount of angular momentum is present and this consideration could imply that the structure formed according to our scenario mainly concerns galaxies associated with a significant rotation, especially spiral galaxies \cite{spiral1,spiral2,spiral3}. However, if the galaxies correspond, as proposed in the numerical simulations of the subsection below, to one of the two fundamental scales of a condensation phenomenon \cite{kraichnan80}, the vorticity contribution is not 
so large and the highly rotating eddies are significantly affected by viscosity dissipation.

\subsection{Numerical Analysis}
In order to deal with a dimensionless equation, we introduce a fundamental fluid speed $u$ and a fundamental length $L$. We can, thus, define the following dimensionless quantities:
\begin{equation}
\Psi \equiv \frac{\psi}{u}\,\frac{2\pi}{L}\, ,\qquad
\bar{x}\equiv x\,\frac{2\pi}{L}\, ,\qquad
\bar{y}\equiv y\,\frac{2\pi}{L}\, ,\qquad
\tau \equiv ut\,\frac{2\pi}{L}\, ,\qquad
\bar{\nu} \equiv \frac{\nu}{u}\,\frac{2\pi}{L}\, .
\label{xxxjea}
\end{equation}
The 
Euler Equation (\ref{jea1x7}) can be now restated as
\begin{equation}
\partial_{\tau}\mathcal{D}^2\Psi +
\partial_{\bar{x}}\Psi\partial_{\bar{y}}\mathcal{D}^2\Psi -
\partial_{\bar{y}}\Psi\partial_{\bar{x}}\mathcal{D}^2\Psi = \bar{\nu}
\mathcal{D}^4\Psi\, ,
\label{xxxjea2}
\end{equation}
where $\mathcal{D}^2$ denotes the Laplacian in the barred coordinates. 

Let us now assume periodic boundary conditions, which allow us to numerically simulate the equation above by means of a truncated Fourier approach. Specifically, we consider the Fourier expansion $
\Psi(\bar{x},\bar{y},\tau) = \sum_{\ell,m}\bar{\Psi}_{\ell,m}(\tau)\;e^{i(\ell \bar{x}+m\bar{y})}$. Here, the mode numbers $(l,m)$ are integers and the reality constraint reads $\bar{\Psi}_{-\ell,-m} =\bar{\Psi}_{\ell,m}^*$. When addressing the spectral evolution, the non-linear terms in Equation (\ref{xxxjea2}) are treated by implementing a pseudo-spectral approach for the resulting convolution product, while a fourth-order Runge--Kutta algorithm evolves the components $\bar{\Psi}_{\ell,m}(\tau)$ in time. In the following, we set $u=300$ km/s 
(according to a typical value in spiral galaxies) and $L=0.01$ Mpc (which is the present day spatial scale of a spiral galaxy). For the viscosity parameter, we used a hydrogen-like plasma with $T=50$ K and $\rho=10^{-26}$ g cm$^{-3}$. We thus obtained, from tabulated data, \mbox{$\eta\simeq1.8\times10^{-5}$ g cm$^{-1}$ s$^{-1}$}, and thus, $\bar{\nu}\simeq2\times10^{-8}$. 

\begin{figure}[ht!]
\centering
\includegraphics[width=9cm]{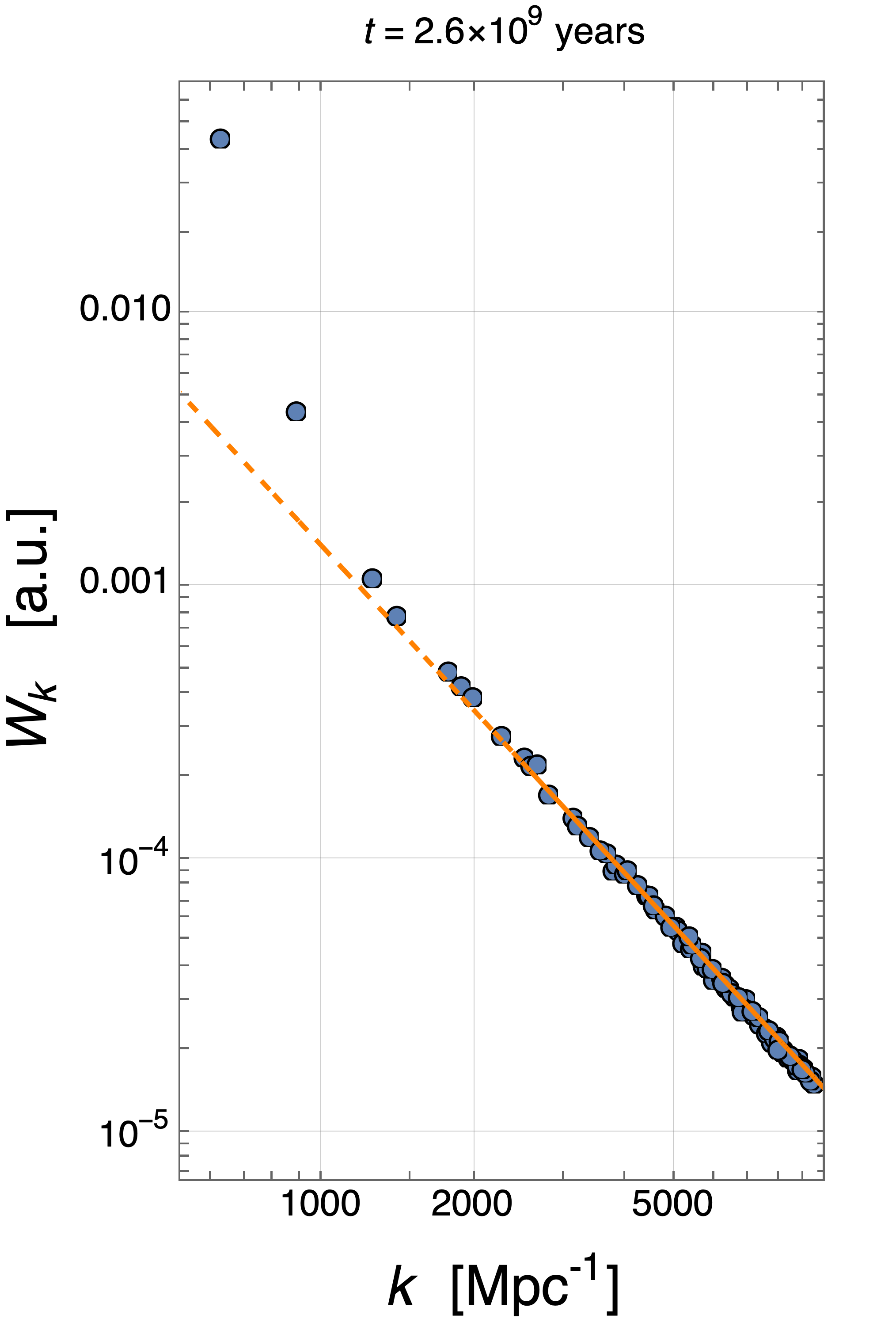}
\caption{Log$-$log 
 plot of $W_k$ (blue circles) 
 in arbitrary units as a function of $k$ at a fixed time, as indicated above the graph. The orange dashed line indicates the behavior $\propto 1/k^2$. 
 \label{figspctr}}
\end{figure}

The energy spectrum was obtained using the quantity $W_k=k^2|\bar{\Psi}_{\ell,m}|^2$, where the definition $k^2=(2\pi/L)^2(\ell^2+m^2)$ held and averages over equal $(\ell^2+m^2)$ values were implemented. 
In Figure \ref{figspctr}, 
we plot $W_k$ at a fixed time evolved from an initial
condition in which three modes were non-zero and their wavenumbers were taken
in the ratio corresponding to $N=2$, i.e., $k=1,\,2,\,4$. We clearly see that the energy was mainly concentrated in the first two modes, while the remaining part of the spectrum rapidly decayed as
$1/k^2$. This was the result of the so-called condensation phenomenon, which is
well-known in fluid dynamics \cite{kraichnan80}: while a direct cascade of
enstrophy was observable and it corresponded to the $1/k^2$ profile (see \cite{montani-fluids2022} for a detailed discussion), an inverse cascade
of energy produced the lowest $k$ values. This scenario, when referred to the cosmological setting, clarifies why we
reliably apply the fragmentation
due to eddy vorticity to
two spatial scales only: de facto clusters of galaxies and galaxies.
In other words, we propose the idea that the fractal turbulence of the
baryon fluid can play a role in
structure formation, but its
fingerprint can concern the formation of galaxies only and, in this sense,
observational evidence of fractality could be recovered at this spatial scale of the Universe.

From the perspective of comparing the idea
proposed here (where a full development requires significant effort that involves
theoretical and numerical investigation) with observational data, the most promising prediction is to be identified in Equation (\ref{jea23}). A statistical analysis of the cluster and galaxy distribution in space
can offer a suitable sample of data
to be compared with Figure \ref{fig1}. In this respect, the incoming data sets from the James Webb Space Telescope and Euclid could provide the required information.

\section{Concluding Remarks}
Here, we present a conceptual framework for interpreting the evolutionary step linking clusters of galaxies to individual galaxies as a consequence of a fragmentation process within turbulent eddies following a fractal distribution law \cite{landau-v06,peebles}.  
The specific aspect in which the turbulence of the cosmological fluid within the non-linear regime influences the gravitational stability of overdensities was identified as the coincidence between the eddy scale and the Jeans length during that stage of the Universe evolution.

It is important to emphasize that when discussing galaxy clusters fragmenting into galaxies, our focus was clearly on proto-structures of this nature, which people commonly refer to as ``non-linear seeds'' in the structure formation. In other words, the concept developed above illustrates the formation of proto-galaxies as sub-eddies within the original turbulent eddy corresponding to a single cluster. The crucial point here is that as a residual influence of the system linear stability (as per the so-called Jeans fragmentation mechanism \cite{ruffini}), 
the formation of galaxies is dictated by the scale corresponding to their Jeans length during that primordial cosmological evolution.

The primary merit of this analysis lay in determining the phenomenological law in Equation (\ref{jea23}), which provides a clear guideline for the conjectured fragmentation mechanism when compared against statistically representative data samples. By evaluating the density transitioning from a cluster to an average contained galaxy, and considering the number of galaxies within the cluster, we can estimate the Hausdorff number for the given large-scale matter distribution. Through the analysis of statistical samples, we could approximate the average Hausdorff number across the Universe's large-scale matter distribution. Consequently, this proposed conjecture serves as a valuable tool for assessing the tendency of the Universe's large-scale structures to exhibit certain fractal features, which based on a rough estimate, seem to be present in some cases.

The introduction of the magnetic field in the primordial Universe
allowed us to clarify how the turbulent dynamics characterizing the
perturbation non-linear growth could be reduced to a 2D Euler
equation. In fact, although the plasma component of the Universe is, after
the recombination age, a very small fraction ($\simeq$2.5 $\times~10^{-4}$), the strong ambipolar diffusion at the spatial scales of
cosmological interest induces a coupling with the neutral hydrogen fluid,
which has to be taken into account in the structure formation problem.
Since the magnetic field provides an anisotropic pressure contribution to
the linear plasma perturbations, this squeezing effect onto the plane
perpendicular to the magnetic field is
transferred to the neutral hydrogen.
We were then led to argue that the fully non-linear dynamics of the baryon
fluctuations is governed by an essentially 2D scheme,
to which we could also apply the Boussinesq approximation as a consequence
of the high value of the plasma $\beta$-parameter. The numerical simulations we performed were initialized by considering only
three major modes, where the modulus
corresponds to a geometrical progression of factor $2$. We demonstrated that a continuous
energy spectrum emerges with a strong trace of an inverse cascade of
energy. In fact, only the first two modes (for us, this corresponded to the
seeds of clusters and galaxies, respectively) have a significant
energy content. The amplitude of the
remaining spectrum decays as $1/k^2$ according to an enstrophy direct
cascade. This feature, which corresponds to a condensation of the spectrum,
can explain why the process of structure formation is associated with a
turbulent fragmentation for what concerns only a few (de facto two)
spatial scales. Moreover, the fractal nature
of the eddy fragmentation can be recovered at the smallest (galactic)
scale of matter distribution.

\section*{Appendix A}
In this appendix, we study in some detail the dispersion relation in Equation (\ref{jea10}). We already stressed that when taking a solution as $\omega=\omega_r+i\gamma$, the resulting values of $\gamma$ are always negative, thus yielding damped and stable behavior of the density contrast. In what follows, we instead discuss the specific solution branch with a pure imaginary frequency, i.e., $\omega=i\gamma$. The dispersion relation is rewritten as
\begin{align}\label{jea-a1}
\gamma^2+\mu k^2 \gamma +k^2 v_s^2 - 4\pi G \rho_0=0\;.
\end{align}
The 
solutions of the equation above constitute two branches: the negative one again results in a stable evolution, and thus, we focused on the following solution:
{\small \begin{align}\label{jea-a2}
\gamma=-\frac{\mu k^2}{2} +\frac{1}{2}\sqrt{\mu^2k^4-4v_s^2k^2+16\pi G\rho_0}\;.    
\end{align}}
Considering 
the stability constraint $\gamma\leqslant0$, the viscous term cancels out and it provides the standard Jeans condition in Equation (\ref{jea11}), i.e., $k\geqslant k_J$, where
\begin{align}\label{jea-a3}
k_J=\sqrt{\frac{4\pi G\rho_0}{v_s^2}}\;.
\end{align}

This 
analysis enforces the study presented in this work for which the inertial regime was assumed by neglecting viscosity effects in determining the Jeans length. However, attention must be put on the compatibility conditions of the solution. In particular, the inequality
\begin{align}\label{jea-a4}
\mu^2k^4-4v_s^2k^2+16\pi G\rho_0\geqslant0\;,    
\end{align}
must hold. Such a relationship provides an upper bound for the viscous parameter derived from the reality condition of the solutions, namely,
\begin{align}\label{jea-a5}
\mu\leqslant\frac{v_s^2}{\sqrt{4\pi G\rho_0}}\;,
\end{align}
which indicates how the viscosity is strongly constrained when the density is assumed to be time dependent and starts to grow. Following Equation (\ref{jea-a5}), we can define the new parameter $u$ as 
\begin{align}\label{jea-a6}
u=\frac{k_J}{v_s/\mu}\;,\qquad0<u\leqslant 1\;,
\end{align}
where we excluded the null value since it should correspond to $\mu=0$ and the whole analysis should fall into the standard Jeans condition. In this scheme, Equation (\ref{jea-a4}) is satisfied for 
\begin{align}\label{jea-a7}
k\leqslant k_-\;,\qquad  k_+\leqslant k\;,\qquad
k_\pm=\frac{k_J}{u}\sqrt{2\Big(1\pm\sqrt{1-u^2}\Big)}\;,
\end{align}
and we have $k_-\leqslant k_+$ by construction. We can deduce how a compatibility gap emerges in the $k$ space if $u\neq1$; in fact, the $\gamma$ solution in Equation (\ref{jea-a2}) is not valid for $k_-<k<k_+$.

\begin{figure}[ht!]
\centering
\includegraphics[width=10cm]{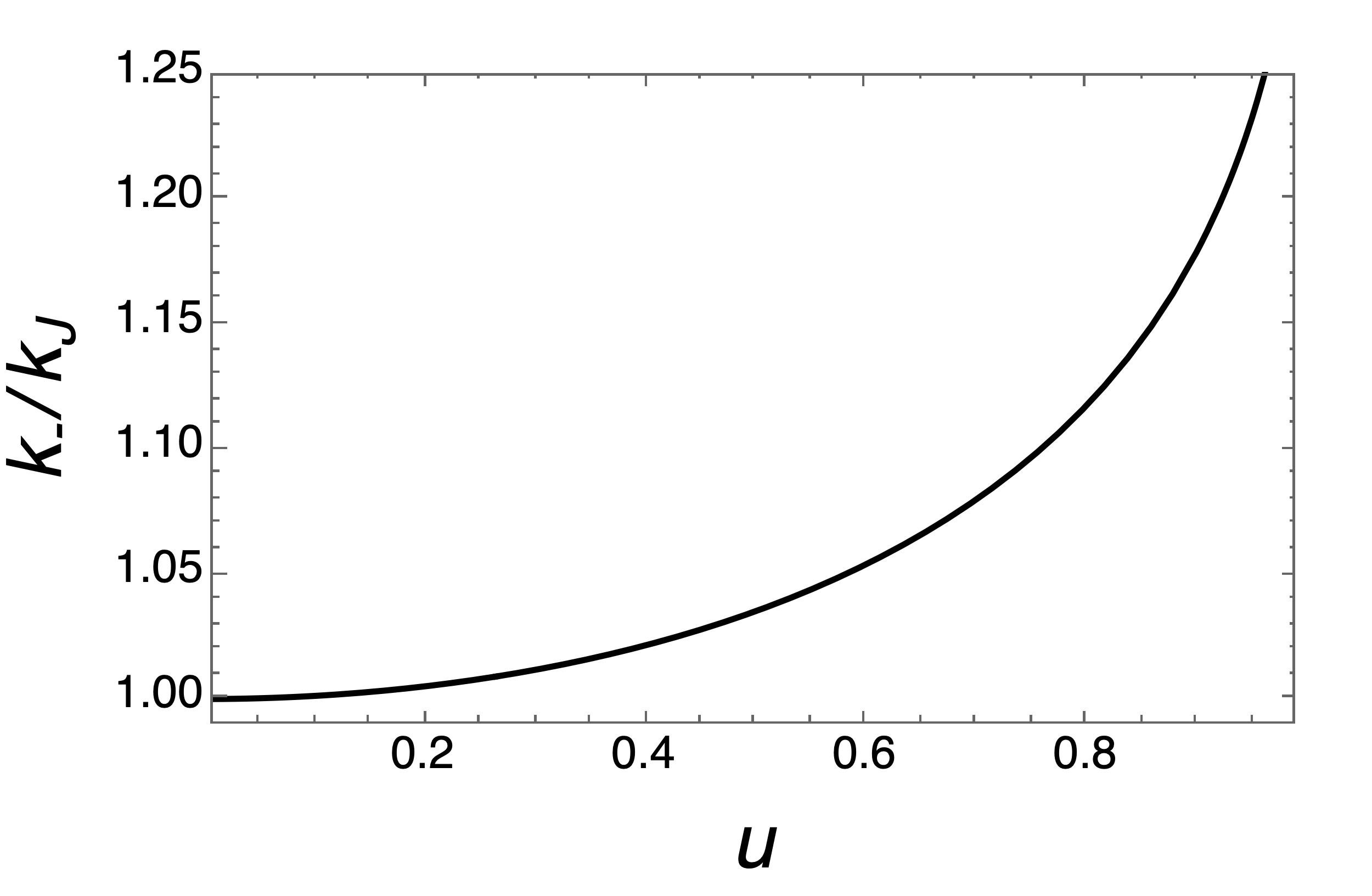}
\caption{Plot of $k_-/k_J$ from Equation (\ref{jea-a7}) as a function of $u$.
\label{fig2}}
\end{figure}

At the same time, when the viscous term is maximized according to Equation (\ref{jea-a5}), thus implying $u=1$, we obtain $k_-=k_+$ and the all the $k$ values are allowed. We recall that these are compatibility conditions and not stability constraints: the density contrast evolution is stable according to the Jeans condition provided in Equation (\ref{jea-a3}). In this respect, we now compare the critical values $k_\pm$ to $k_J$ when $u\neq1$. Provided that in this range, $k_-< k_+$, in Figure \ref{fig2}, it can be appreciated how $k_J<k_-$. This demonstrates the existence of an available region of gravitational instability of our system, even in the presence of a viscous contribution.

\vspace{-6pt}

\section*{Appendix B}
In this appendix, we derive the dispersion relation in Equation (\ref{jea1x}). The dynamics of the magnetized cosmological plasma is described by the continuity Equation (\ref{jea1}), the Poisson Equation (\ref{jea3}), and the Euler Equation (\ref{jea2}) corrected by the presence of a magnetic field $\textbf{B}$. Such a system must be coupled with the evolution equation for $\textbf{B}$ and by the Gauss condition $\boldsymbol{\nabla}\cdot\textbf{B}=0$. In the case under consideration, we neglect the viscous terms in the Euler equation for the sake of simplicity (see \cite{montani-incarbone} for a detailed analysis). The complete set of dynamical equations thus reads
\begin{align}
\partial_t\rho + \boldsymbol{\nabla}\cdot (\rho \textbf{v}) = 0\, ,\\
\Delta \Phi -4\pi G \rho=0\,,\\
\rho \left( \partial _t\textbf{v}  + \textbf{v}\cdot \boldsymbol{\nabla}\textbf{v}\right) +\boldsymbol{\nabla}p + \rho \boldsymbol{\nabla}\Phi -(\boldsymbol{\nabla}\times\textbf{B})\times\textbf{B}/4\pi=0 \,,\\
\partial_t\textbf{B}-\boldsymbol{\nabla}\times(\textbf{v}\times\textbf{B})=0\,.
\end{align}

As  
in Section \ref{sec2}, we address a static background characterized by constant density and magnetic field and by a null velocity and gravitational potential (for these two quantities, we retain the same notation). Perturbing the equations above up to first order under such conditions, the system can be reduced to the following relation for the velocity fluctuations:
\begin{align}
    \partial_t^2\textbf{v}-v_s^2 \boldsymbol{\nabla}(\boldsymbol{\nabla}\cdot\textbf{v})+\boldsymbol{\nabla}\partial_t\Phi+v_A^2\hat{\textbf{B}}_0\times(\hat{\textbf{B}}_0\times(\boldsymbol{\nabla}(\boldsymbol{\nabla}\cdot\textbf{v})))+\nonumber
\qquad\qquad\\   +v_A^2\hat{\textbf{B}}_0\times((\hat{\textbf{B}}_0\cdot\boldsymbol{\nabla})(\boldsymbol{\nabla}\times\textbf{v}))=0\,,
\label{appbv1}
\end{align}
where, as already discussed, $v_A$ indicates the background Alfv\'en velocity calculated as $v_A^2\equiv B_0^2/4\pi \rho_b$. Introducing the plane wave expansion $\textbf{v}\propto e^{i\textbf{k}\cdot\textbf{x}}$ and decomposing the velocity field as $\textbf{v}=v_{\parallel} \textbf{k}+\textbf{v}_\perp$, Equation (\ref{appbv1}) splits into the system
\begin{align}
\partial_t^2 v_{\parallel}+(k^2\tilde{v}^2+(1-\cos^2\theta)k^2 v_A^2)v_{\parallel}-k^2 v_A^2\cos\theta(\hat{\textbf{B}}_0\cdot\textbf{v}_\perp)=0\,,\\
\partial_t^2\textbf{v}_\perp+k^2v_A^2 \textbf{v}_\perp\cos\theta+
k^2 v_A^2\cos\theta(\hat{\textbf{k}}\cos\theta-\hat{\textbf{B}}_0)=0\,,
\end{align}
where $\tilde{v}^2\equiv v_s^2 - 4\pi G \rho_0/k^2$ and $\cos \theta \equiv 
\hat{\textbf{B}}_0\cdot\hat{\textbf{k}}$. The system above reduces to a single  equation for the field $v_\parallel$, and taking $v_\parallel\propto e^{i\omega t}$, we finally obtain the dispersion relation written as
\begin{align}
\omega^4-k^2(\tilde{v}^2+v_A^2)\omega^2+k^4\tilde{v}^2v_A^2\cos^2\theta=0\,,
\end{align}
admitting the solutions in Equation (\ref{jea1x}).


\end{document}